\documentclass[final,3p,times]{elsarticle}

\usepackage{amssymb}
\usepackage{amsmath}

\journal{arXiv}

\begin{document}

\begin{frontmatter}



\title{Maximum mass of singularity-free anisotropic compact stars in Rastall theory of gravity}


\author{Sourav Biswas, Debadri Bhattacharjee \& Pradip Kumar Chattopadhyay} 

\address{IUCAA Centre for Astronomy Research and Development (ICARD), Department of Physics, Cooch Behar Panchanan Barma University, Vivekananda Street, District: Cooch Behar,  Pin: 736101, West Bengal, India.}
\ead{souravcbpbu@gmail.com, debadriwork@gmail.com \& pkc$_{-}$76@rediffmail.com}

\begin{abstract}
	The current model explores spherically symmetric anisotropic compact stars within the Rastall theory of gravity. By employing the Krori and Barua metric ansatz (K.D. Krori and J. Barua, J. Phys. A: Math. Gen. 8 (1975) 508), we derive a set of tractable, singularity-free relativistic solutions to the Einstein field equations. Using a best-fit equation for the numerical solution of the TOV equation, we determine the maximum mass and corresponding radius in this model. Our findings reveal that an increase in the Rastall parameter $(\xi)$ leads to a higher maximum mass, indicating a stiffer nature of the equation of state. For $\xi$ values ranging from 0.01 to 0.09, we calculate the maximum mass to be between $2.24M_{\odot}$ and $2.36M_{\odot}$, with corresponding radii from 9.48 to 10.15 km. Furthermore, our model's predictions for the radii of recently observed pulsars are consistent with observational data. The model satisfies essential criteria for causality, energy conditions, and stability, confirming its viability and physical acceptability as a stellar structure.
\end{abstract}




\end{frontmatter}
\section{Introduction}\label{sec1}
Albert Einstein's formulation of General Relativity (GR) marked the inception of a new era in relativistic astrophysics. GR has established itself as a theory of gravity with great success, and since then, extensive efforts have been devoted to construct stellar models in the framework of GR. Observational investigations on gravitational redshift, shapiro delay, bending of light beam due to the effects of gravitational lensing and precession of perihelion of mercury in solar system \cite{Will} affirmed the validity of GR precisely. Furthermore, the detection of gravitational waves from GW events, such as, binary black hole mergers and neutron star (NS) mergers \cite{Stairs,Abbott,Abbott1} by the advanced LIGO and VIRGO detectors has provided additional confirmation of the robustness of GR outside the Solar System. These observations allow for testing GR in strong gravity regimes, as the merger of two black holes involves intense gravitational fields \cite{Abbott1}. However, it remains uncertain whether GR is valid in this extreme regime. The apparent failure of the equivalence principle at a black hole's singularity challenges the concept of GR's universality.

Despite GR's remarkable success in predicting a wide range of gravitational phenomena and recent gravitational wave observations, it encounters significant theoretical challenges when addressing phenomena from the accelerating universe to the cosmic dark sector across various scales. Thus, identifying necessary corrections to GR from the perspective of both classical and quantum approaches in weak as well as strong gravity regimes is essential. In such context, modified gravity theories have provided valuable insights. Interest in the necessary modifications to Einstein's gravity, theoretically, acts as a strong motivation to the scientific community. Additionally, recent observational evidences support the exploration of such type of modifications. For extensive understanding of the cosmos, pursuing gravitational theories in alternative forms is crucial although fundamental framework for such studies remains to be GR. 

A common approach to modifying GR involves altering the action of Einstein-Hilbert, which comprises of the Lagrangian density function $\sqrt{-g}R$, where $R$ is termed Ricci scalar of the curved space-time. The necessity for modified gravity theories is extensively discussed in \cite{Nojiri}. Consequently, researchers have generalised the Einstein-Hilbert action to formulate various modified gravity theories \cite{Capozziello,Elizalde,Bamba,Houndjo,Yousaf,Ilyas} such as $f(R)$, $f(T)$, $f(R,T)$, $f(G)$, and $f(Q)$ etc. Many researchers have explored these theories to address different aspects of astrophysics and cosmology \cite{Das,Das1,Das2,Biswas,Lohakare,Xu,Doneva,Yazadjiev,Feola}. These theories often hinge on the concept of the non-conservation principle of the energy-momentum tensor. P. Rastall proposed a new modification of GR, where it is considered that in the curved space-time, the covariant divergence is non-vanishing, and is known as Rastall gravity \cite{Rastall,Rastall1}. The covariant formulation of the divergence of energy momentum tensor $T_{\mu\nu}$ is expressed as $\nabla_{\nu} T_{\mu}^{\nu}=\gamma R_{,\mu}$, where $R_{,\mu}$ is the covariant derivative of $R_{\mu\nu}$. 

The Rastall theory provides a more concise as well as manageable form of the Einstein field equations (EFE), yielding fascinating consequences on a cosmic scale. Recently, Oliveira et al. \cite{Oliveira} explored the neutron star model within the framework of Rastall gravity. Additionally, W.E. Hanafy \cite{Hanafy} conducted an interesting study on the effects of Rastall gravity on the gross properties of PSR J0740+6620. Hansraj et al. \cite{Hansraj} investigated the impact of Rastall parameter on the perfect fluid distribution in a spherically symmetric spacetime. Various aspects of astrophysics and cosmology within the Rastall theory have been examined by several researchers, including Heydarzade et al. \cite{Heydarzade}, Parihadi et al. \cite{Parihadi}, Kumar and Ghosh \cite{Kumar}, Xu and Wang \cite{Xu1}, Ma and Zhao \cite{Ma}, Gergess and Nashed \cite{Gergess}, Mota et al. \cite{Mota}, Abbas and Shahzad \cite{Abbas}, among others. 

Observational data from stellar bodies have paved the way for developing new theories to understand stellar properties. Compact objects, representing the final stages of stellar evolution, offer theoretical understanding of highly dense matter configurations. Over the past decade, advancements in theoretical modeling have facilitated the transition from theoretical perspectives to analytical solutions. Recently, extensive data from pulsars and gravitational wave (GW) events have been analysed by many researchers, leading to accurate estimations of numerous physical parameters of these astrophysical objects.

One notable achievement was the precise mass determination of the millisecond pulsar PSR J0740+6620 using the Shapiro delay, which yielded a mass of $(2.14)^{+0.10}_{-0.09}~M{\odot}$ \cite{Cromartie}. This has spurred discussions on the maximum mass of compact objects, with suggestions that some compact objects might exceed the mass of PSR J0740+6620 \cite{Cromartie} within interacting systems. The gravitational wave event GW190814 revealed the existence of a companion star having mass $2.59^{+0.12}_{-0.09}~M_{\odot}$ at 90\% confidence \cite{Abbott2}. However, it remains unclear whether this objects are light black holes or a very massive NS or other exotic type of stars. If the later is true, new theoretical concepts would be necessary to extend the upper limit of maximum mass to accommodate such high values. This is very crucial for studying the internal composition of matter above the saturation density.

In this context, anisotropy in pressure is pivotal in determining the maximum possible mass and radius of such compact stars, as the maximum mass as well as their stability both increase with the increase of anisotropy. Ruderman \cite{Ruderman} and Canuto \cite{Canuto}, proposed that anisotropy can emerge locally in the regime of high-density $(\rho>10^{15} gm/{cm}^3)$ and significantly influence the gross properties of matter, interior to such compact stars. Various natural phenomena are held responsible for the possible origin of anisotropy in pressure in highly dense stars which are fermionic fields, electromagnetic fields in NS \cite{Sawyer}, pion condensation \cite{Sawyer1}, and superfluidity \cite{Carter} etc. The review article \cite{Mardan} provides an in-depth investigation into the origins of local anisotropy, establishing viscosity as a potential source \cite{Herrera1}.

Bowers and Liang \cite{Bowers} introduced the concept of pressure anisotropy, exploring its impact on the properties of stellar parameters such as the compactness and mass-radius relationships, surface redshift etc., to generalise the stellar models. Heintzmann and Hillebrandt \cite{Heintzmann} examined the properties of anisotropic NS in relativistic limit, concluding that, in principle, there are no limits to mass or redshift for arbitrarily large anisotropic factors. Following the pioneering work of Carter and Langlois \cite{Carter}, numerous studies \cite{Maurya1}-\cite{Hernandez} have focused on spherically symmetric anisotropic stellar configurations in static equilibrium. Additionally, Kalam et al. \cite{Kalam1} demonstrated how the central density depends on the value anisotropy parameter.

Modeling of spherically symmetric and anisotropic stars provides a more generalised approach to employing the equation of state (EoS) of interior matter. Tolman-Oppenheimer-Volkoff (TOV) equation \cite{Tolman}-\cite{Oppenheimer} can be distinctively used to figure out the interior structure of stellar configurations as well as their mass-radius relation. Moreover, it must be acknowledged that theoretical models are greatly influenced by the choice of metric potentials used to derive new classes of exact relativistic solutions to the EFE. The Krori-Barua (KB) metric ansatz \cite{KB} is among the most recognised approaches for describing a spherically symmetric perfect fluid distribution. The singularity-free solutions generated by this method are pivotal in relativistic stellar modeling. Numerous studies have utilised the KB metric ansatz for stellar modeling \cite{Varela, Kalam2, Hossein, Rahaman2, Bhar3}. 

In the present article, we focus on the theoretical study of the structural and dynamical properties of anisotropic compact stars within the framework of Rastall theory of gravity. Our analysis indicates that Rastall gravity offers a potential alternative framework to GR, capable of addressing certain theoretical limitations. The modified field equations within this theory may accommodate solutions unattainable in GR, potentially leading to a broader class of compact star models with enhanced stability and distinct physical properties. Comparative studies of compact stellar structures in both Rastall gravity and GR can serve as a critical test of the robustness of GR and its limitations. Observational evidence favouring Rastall gravity over GR would necessitate a paradigm shift in our understanding of gravity. A key finding of the present work is that Rastall gravity affords greater flexibility in stellar modeling, enabling the exploration of a wider range of configurations. Such flexibility may prove the usefulness of Rastall theory of gravity in explaining the challenging phenomena of astrophysics, which are, otherwise, difficult to explain within the GR paradigm. Furthermore, anisotropic models within the context of Rastall gravity may exhibit unique observational signatures, such as distinct mass-radius relation, tidal deformabilities and gravitational wave patterns, offering potential avenues for discriminating between the two theories through astrophysical observations.

The paper is organised as follows: Section \ref{sec2} addresses the solutions of the modified EFE within the Rastall theory of gravity. In Section \ref{sec3}, we match the interior solutions with the exterior one through the extrinsic curvature tensors to obtain the necessary parameters present in the model. The physical application of the model described in Section \ref{sec4}. In this section, we demonstrate the energy density and pressure profiles, anisotropy parameter, along with the study of physical viability of the model through causality and energy conditions through graphical representation. Section \ref{sec5} describes the stability analysis of model on the basis of generalised TOV equation, Herrera's cracking concept and the radial variation of Adiabatic index. Finally, we summarise the major findings of the paper in Section \ref{sec6}. 
\section{Einstein field equations in Rastall theory of gravity and their solutions}\label{sec2}
P. Rastall \cite{Rastall,Rastall1} proposed the concept of a non-vanishing divergence of $T_{\mu}^{\nu}$, which is known as the matter energy-momentum tensor, that subsequently led to modifications in Einstein's gravity. In the Rastall formalism, the divergence of $T_{\mu}^{\nu}$ is directly proportional to the covariant derivative of the Ricci scalar and it is expressed as:
\begin{equation}
	\nabla_{\nu} T_{\mu}^{\nu}=\zeta_{\mu}, \label{eq1}
\end{equation}
where,
\begin{equation}
	\zeta_{\mu}=\gamma R_{,\mu}, \label{eq2}
\end{equation}
where, $\gamma$ is a constant, the Ricci scalar is demarcated as $R=g^{\mu\nu} R_{\mu \nu}$ and $T_{\mu}^{\nu}$ describes the energy-momentum tensor. In this background of Rastall modifications, the EFE takes the form:
\begin{equation}
	G_{\mu\nu}+k\gamma g_{\mu\nu}R =kT_{\mu\nu}. \label{eq3}
\end{equation}
In the fundamental description of the modified form of EFE in Rastall theory, the Ricci tensor $R_{\mu \nu}$, Ricci scalar $(R)$ and $T_{\mu \nu}$, the energy-momentum tensor are related via the following relation:
\begin{equation}
	R_{\mu \nu}+ (k\gamma-\frac{1}{2})g_{\mu\nu}R=kT_{\mu\nu}. \label{eq4}
\end{equation}
Here, $\lq k$\rq~represents the gravitational coupling constant $(=\frac{8\pi G}{c^{4}})$. Now, contracting Eq.~\ref{eq4}, we obtain,
\begin{equation}
	R(4k\gamma-1)=kT_{\mu}^{\mu}. \label{eq5}
\end{equation}
Interestingly, in Eq.~\ref{eq5}, for the choice of $k\gamma=\frac{1}{4}$, we obtain, $T_{\mu}^{\mu}=0$ and $R=0$ which in turn leads to the vacuum solutions of EFE with $R_{\mu\nu}=0$. Therefore, the case of $k\gamma=\frac{1}{4}$ is excluded. For simplification, we consider $\xi$=k$\gamma$ as the dimensionless Rastall parameter. Now, in the Rastall theory of gravity, Rastall parameter $(\xi)$ and coupling constant $(k)$ take the forms:
\begin{equation}
	k=\frac{8\pi G}{c^{4}}\Bigg(\frac{4\xi-1}{6\xi-1}\Bigg), \label{eq6}
\end{equation}
and 
\begin{equation}
	\gamma=\frac{c^{4}\xi}{8\pi G}\Bigg(\frac{6\xi-1}{4\xi-1}\Bigg). \label{eq7}
\end{equation}
Notably, substituting $\xi$=0 in Eq.~\ref{eq6} returns the value of $k$ in GR. Again, from Eq.~\ref{eq6} and Eq.~\ref{eq7}, it is evident that $k$ and $\gamma$ show a diverging nature for $\xi=\frac{1}{6}$ and $\xi=\frac{1}{4}$ respectively. Therefore, in this framework, we exclude $\xi=\frac{1}{6}$ and $\xi=\frac{1}{4}$. In the relativistic system of units $(G=1,~c=1)$, the EFE in the framework of Rastall theory takes the form:
\begin{equation}
	G_{\mu\nu}+\xi g_{\mu\nu}R=8\pi T_{\mu\nu} \Bigg(\frac{4\xi-1}{6\xi-1}\Bigg).\label{eq8}
\end{equation}
In the spherically symmetric space-time coordinates, $X^{\mu}=(t,r,\theta,\phi)$, we consider the form of the static line element as:
\begin{equation}
	ds^2=-e^{2\nu(r)}dt^2+e^{2\lambda(r)}dr^2+r^2d\theta^2+r^2sin^2\theta d\phi^2. \label{eq9}
\end{equation}
The energy-momentum tensor $(T_{\mu\nu})$ for a perfect fluid distribution is expressed as:
\begin{equation}
	T_{\mu\nu}=diag(-\rho,p_{r},p_{t},p_{t}) \label{eq10}
\end{equation}
here, the terms $\rho=$, $p_{r}=$ and $p_{t}=$ represent respectively, the energy density, radial pressure and tangential pressure. By using Eq.~\ref{eq8} and Eq.~\ref{eq9}, in Eq.~\ref{eq10}, we get the tractable set of following equations:
\begin{equation}
	\hspace{-3cm}
	\frac{2\lambda'e^{-2\lambda}}{r} +\frac{(1-e^{-2\lambda}}{r^2}+\xi e^{-2\lambda}\Bigg[2
	\nu''+2(\nu')^2-2\lambda'\nu'+\frac{4(\nu'-\lambda')}{r}+\frac{2(1-e^{2\lambda})}{r^2}\Bigg] =8\pi\rho\Bigg(\frac{4\xi-1}{6\xi-1}\Bigg),\label{eq11}
\end{equation}
\begin{equation}
	\hspace{-3cm}
	\frac{2\nu'e^{-2\lambda}}{r}-\frac{1-e^{-2\lambda}}{r^2}-\xi e^{-2\lambda}\Bigg[2
	\nu''+2(\nu')^2-2\lambda'\nu'+\frac{4(\nu'-\lambda')}{r}+\frac{2(1-e^{2\lambda})}{r^2}\Bigg]=8\pi p_{r}\Bigg(\frac{4\xi-1}{6\xi-1}\Bigg),\label{eq12}
\end{equation}
\begin{equation}
	\hspace{-3.5cm}
	e^{-2\lambda}\Bigg[\nu''+(\nu')^2-\lambda'\nu'+\frac{\nu'-\lambda'}{r}\Bigg]-\xi e^{-2\lambda}\Bigg[2
	\nu''+2(\nu')^2-2\lambda'\nu'+\frac{4(\nu'-\lambda')}{r}+\frac{2(1-e^{2\lambda})}{r^2}\Bigg]=8\pi p_{t}\Bigg(\frac{4\xi-1}{6\xi-1}\Bigg).\label{eq13}
\end{equation}	 
where, the overhead prime $(')$ on $\lambda$ and $\nu$ represents differentiation with respect to r. To solve the system of equations and to obtain a singularity free solution, we consider metric ansatz as proposed by Krori-Barua (KB) \cite{KB} as, $\lambda=ar^2$ and $\nu=br^2+c$, where a, b and c are arbitrary constants and a and b have the dimensions of $Km^{-2}$ and c is dimensionless. Now, substituting the metric ansatz in Eq.~\ref{eq11}, Eq.~\ref{eq12} and Eq.~\ref{eq13}, we obtain the analytical expression of $\rho,~p_{r}$ and $p_{t}$ as:
\begin{equation}
	\hspace{-2.5cm}
	\rho=-\frac{e^{-2ar^2}(6\xi-1)}{8\pi r^2(4\xi-1)}\Bigg(1-e^{2ar^2}-4ar^2-2\xi+2e^{2ar^2}\xi+8ar^2\xi-12br^2\xi+8abr^4\xi-8b^2r^4\xi\Bigg), \label{eq14}	
\end{equation}
\begin{equation}
	\hspace{-2.5cm}
	p_{r}=\frac{e^{-2ar^2}(6\xi-1)}{8\pi r^2(4\xi-1)}\Bigg(1-e^{2ar^2}+4br^2-2\xi+2e^{2ar^2}\xi+8ar^2\xi-12br^2\xi+8abr^4\xi-8b^2r^4\xi\Bigg), \label{eq15}	
\end{equation}
\begin{equation}
	\hspace{-3.5cm}
	p_{t}=\frac{e^{-2ar^2}(6\xi-1)}{4\pi r^2(4\xi-1)}\Bigg(-ar^2+2br^2-2abr^4+2b^2r^4-\xi+e^{2ar^2}\xi+4ar^2\xi-6br^2\xi+4abr^4\xi-4b^2r^4\xi\Bigg).\label{eq16}
\end{equation}
The pressure anisotropy ($\Delta$) inside a star is actually the difference of radial $p_{r}$ and tangential $p_{t}$ pressures, and is expressed as,
\begin{equation}
	\Delta=p_{t}-p_{r}  \label{17}
\end{equation}
\section{Boundary condition}\label{sec3}
The space-time manifold $(\Omega)$ in 4-dimensions is divided into two distinct regions: (i) interior and (ii) exterior. These two regions are separated by a hypersurface $(\Sigma)$ of three dimensions characterised by a induced metric $h_{ij}$ in the $X^{i}$ coordinate system, where the indices i,j, exclude the direction which is orthogonal to hypersurface $\Sigma$. The projection tensors as well as the normal vector from $\Omega$ to $\Sigma$ are given by $e^{a}_{i} = \frac{\partial x^{a}}{\partial X^{i}}$ and $n_{a} = \epsilon \partial_{a}\ell$, respectively, where $\ell$ is the affine parameter along the geodesic which is perpendicular to $\Sigma$. The value of $\epsilon$ are -1, 0, or 1 respectively for the geodesics of timelike, null, and spacelike envelopes. By contraction, we have $n^{a} e^{i}_{a}=0$. In this framework, the induced metric $h_{ij}$ as well as the extrinsic curvature $K_{ij}$ of the hypersurface are expressed as follows:
\begin{equation}
	h_{ij}=e^{a}_{i}e^{b}_{j}g_{ab}, ~~~~~~~~~~~~~~K_{ij}=e^{a}_{i}e^{b}_{j}\nabla_{a}n_{b}. \label{eq18}
\end{equation}
At the hypersurface $(\Sigma)$, the induced metric is assumed to be continuous and using the distribution formalism \cite{Rosa}, we derive that $[h_{ij} = 0]$. For a smooth transition, the extrinsic curvature must also be continuous at the surface of the stellar object, represented by $[K_{ij} = 0]$ \cite{Rosa}. To delve deeper into these matching conditions, we need to start by equating the extrinsic curvature tensor $(K_{ij})$ for both the interior and exterior solutions at the surface of the star. The exterior space-time is described by the Schwarzschild solution \cite{Schwarzschild}, which is given by:
\begin{equation}
	ds^{2}=-\Bigg(1-\frac{2M}{r}\Bigg)dt^{2}+\Bigg(1-\frac{2M}{r}\Bigg)^{-1}dr^{2}+r^{2}(d\theta^{2}+sin^{2}\theta d\phi^{2}). \label{eq19}
\end{equation} 
In a spherically symmetric space-time, the extrinsic curvature $K_{ij}^{\pm}$ has only two independent components: $K_{00}$ and $K_{\theta\theta} = K_{\phi\phi}\sin^{2}\theta$. Here, the (+) sign denotes the exterior region whereas (-) sign indicates interior region, respectively. Utilizing the KB ansatz \cite{KB} and Eq.~(\ref{eq19}), we can express these components in the following form:
\begin{equation}
	K_{00}^{+}=-\frac{M}{r^{2}}\sqrt{1-\frac{2M}{r}},~~~~~~~~~~K_{00}^{-}=-2Bre^{-r^{2}(A-2B-2C)}. \label{eq20}
\end{equation}  
and, 
\begin{equation}
	K_{\theta\theta}^{+}=r\sqrt{1-\frac{2M}{r}},~~~~~~~~~~K_{\theta\theta}^{-}=re^{-Ar^{2}}.  \label{eq21}
\end{equation}
Now, matching Eq.~\ref{eq20} and Eq.~\ref{eq21} at the stellar boundary $(r=R)$ and using the condition that at the stellar surface, radial pressure vanishes, i.e., $p_{r}(r=R)=0$, we obtain:
\begin{eqnarray}
	a=-\frac{\log[1-\frac{2M}{R}]}{2R^2} \label{eq22} \\
	b=-\frac{M}{2(2M-R)R^2}  \label{eq23} \\
	c=\frac{1}{2}\Bigg[-2bR^2+\log\Bigg(\frac{M}{2bR^3}\Bigg)\Bigg]\label{eq24}
\end{eqnarray}
\section{Physical application of the model}\label{sec4}
The physical attributes of a compact object within the Rastall gravity are investigated in this section. To assess the model's viability, pulsar PSR J0740+6620, having mass of $2.072~M_{\odot}$ and radius 12.39 Km \cite{Riley}, is employed as a test case. Given the constraints imposed by the Rastall parameter $(\xi)$ within the theory, we consider values of $(\xi)$ equal to 0.01, 0.05, and 0.09 throughout the analysis. To streamline the presentation and circumvent the complexities of analytical expressions of the physical parameters, such as energy density $(\rho)$, pressures along radial $(p_{r})$, and tangential $(p_{t})$ directions, we opt for a graphical representation of their radial profiles, alongside the anisotropy parameter $(\Delta)$, causality condition, and energy conditions.
\subsection{Energy density and pressure profile}
\begin{figure}[h]
	\centering
	\includegraphics[width=10cm]{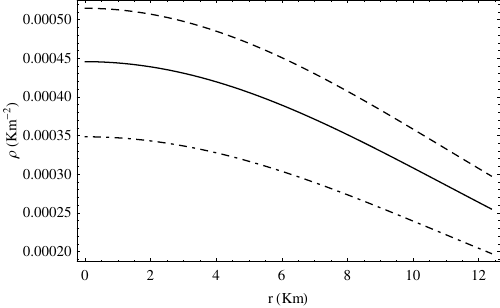}
	\caption{Variation of energy density $\rho$ with radius $r$. Here, Dashed, Solid and Dot-Dashed lines represent $\xi=0.01,~0.05$ and $0.09$ respectively.}
	\label{fig1}
\end{figure}

\begin{figure}[h]
	\centering
	\includegraphics[width=10cm]{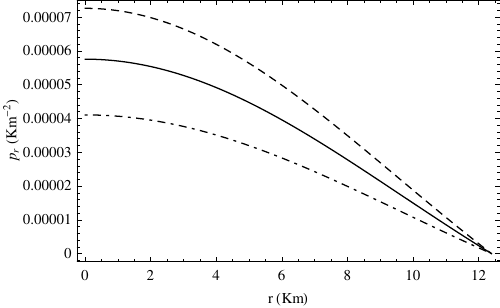}
	\caption{Variation of radial pressure $p_{r}$ with radius $r$. Here, Dashed, Solid and Dot-Dashed lines represent $\xi=0.01,~0.05$ and $0.09$ respectively.}
	\label{fig2}
\end{figure}

\begin{figure}[h]
	\centering
	\includegraphics[width=10cm]{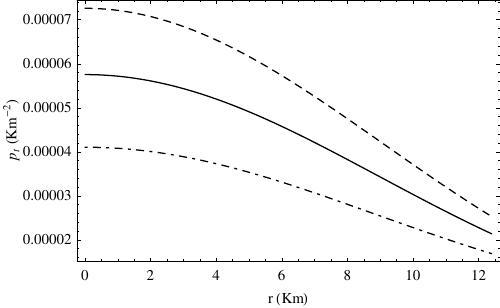}
	\caption{Variation of transverse pressure $p_{t}$ with radius $r$. Here, Dashed, Solid and Dot-Dashed lines represent $\xi=0.01,~0.05$ and $0.09$ respectively.}
	\label{fig3}
\end{figure}

Fig.~\ref{fig1}, Fig.~\ref{fig2} and Fig.~\ref{fig3} illustrate the energy density and pressure profiles as a function of radial coordinate. It is evident from the figures that the energy density $(\rho)$, radial $(p_{r})$ and tangential pressure $(p_{t})$ are all positive as well as pick up finite values within stellar configuration. It is observed that the values of $\rho$, $p_{r}$ and $p_{t}$ are maximum at the centre of the star and decrease in a monotonous nature towards the stellar boundary. Moreover, it is also noted that the energy density and pressure profiles decrease with the increases of Rastall parameter $(\xi)$, which points towards a more stable configuration.  
\subsection{Anisotropy}
\begin{figure}[h!]
	\centering
	\includegraphics[width=10cm]{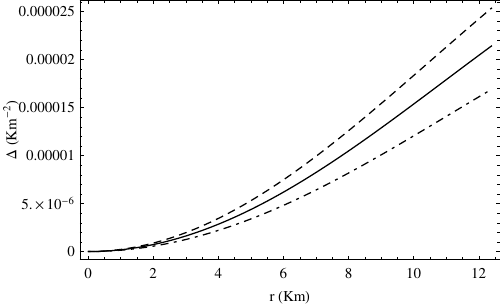}
	\caption{Represantation of anisotropy $\Delta$ with radius $r$. Here, Dashed, Solid and Dot-Dashed lines respectively represent $\xi=0.01,~0.05$ and $0.09$.}
	\label{fig4}
\end{figure}

In this segment, we analyse the graphical response of anisotropy parameter, which is represented by $\Delta$ and defined as the difference between the transverse and radial pressure i.e. $\Delta=p_{t}-p_{r}$. Now, is $\Delta$ is negative $(p_{t}<p_{r})$, the nature of the force will be attractive and in case of repulsive nature, $\Delta$ will be positive $(p_{t}>p_{r})$. From Fig.~\ref{fig4}, it is noted that the anisotropy remains positive through out the stellar interior, which implies that the nature of the force is repulsive, which is necessary to construct an equilibrium configuration of stellar structure \cite{Gokhroo}. 
\subsection{Mass-radius relation from TOV equation} 
\begin{figure}[h!]
	\centering
	\includegraphics[width=10cm]{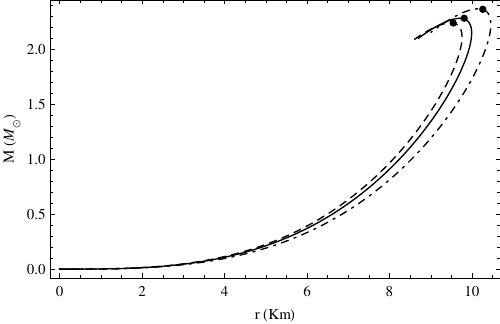}
	\caption{Mass-radius relationship from TOV equation. Here, Dashed, Solid and Dot-Dashed lines respectively represent $\xi=0.01,~0.05$ and $0.09$.}
	\label{fig4a}
\end{figure}

Here, the TOV equations \cite{Tolman,Oppenheimer} is solved to obtain the possible maximum mass and the associated radius in this present model. The numerical solution of TOV equations is entirely based on the choice of particular EoS. In the approach, we have obtained the EoS through the method of sound velocity maximisation at the stellar centre. Now, from the causality condition, we know that sound velocity has an extreme causal limit of $1$ and we employ this apex of causality at the centre, i.e., $v_{r}^{2}|_{r=0}=\frac{dp_{r}}{d\rho}|_{r=0}=1$ along with the method of curve fitting to determine the best fit EoS for different choices of Rastall parameter $(\xi)$. The results are tabulated in Table~\ref{tab1}. 
\begin{table}[h]
	\centering
	\caption{\label{tab1} Best fit EoS for different choices of $\xi$.}
	\vspace{0.5cm}
\begin{tabular}{cc}
			\hline
			Rastall parameter $(\xi)$ & EoS \\  
			\hline
			0.01 & $p_{r}=0.905609\rho-0.000563611$ \\
			0.05 & $p_{r}=0.880296\rho-0.000517796$ \\
			0.09 & $p_{r}=0.844125\rho-0.000447121$ \\
			\hline
		\end{tabular}
\end{table}     
Using Table~\ref{tab1}, we have obtained the numerical solution of the TOV equations to determine the maximum mass and the associated radius in the present model. The results are shown through graphical representation in Fig.~\ref{fig4a}.
\begin{table}[h]
	\centering
	\caption{\label{tab2} Maximum mass and corresponding radius from TOV equations.}
	\vspace{0.5cm}
\begin{tabular}{ccc}
			\hline
			Rastall parameter $(\xi)$ & Maximum mass $(M_{\odot})$ & Radius (Km)\\  
			\hline
			0.01 & 2.24 & 9.48 \\
			0.05 & 2.28 & 9.70 \\
			0.09 & 2.36 & 10.15 \\
			\hline
		\end{tabular}
\end{table} 
The introduction of Rastall theory, allows a non-minimal coupling between matter and gravity. As a result, gravity behaves differently under such extreme conditions in comparison to GR. Now, with increasing Rastall parameter $(\xi)$, we note an effective reduction in pressure relative to the energy density which results in an apparent softer EoS as demonstrated in Table~\ref{tab2}. In GR, a softer EoS, means lower maximum mass. However, Rastall gravity modifies the TOV equation, effectively altering the balance between gravity and pressure. Further with increasing Rastall parameter $(\xi)$, the non-minimal coupling between the matter sector and geometry strengthens which leads to an effective reduction in gravitational strength in higher densities. This effective reduction presents an additional support to the stellar structure to withstand the gravitational collapse. Hence, despite the EoS being softer, the maximum mass and the associated radius increase with increasing Rastall parameter $(\xi)$, as evident from Table~\ref{tab2}. Moreover, in studying the hadronic matter composition of a NS in the framework of Rastall theory of gravity, Pretel \cite{Pretel} obtained similar results regarding the mass-radius relationship. To check for the physical viability of the present model, we have tabulated the predicted radii of some recently observed pulsars along with their central and surface density and central pressure in Table~\ref{tab3} and Table~\ref{tab4}. 
\begin{table}[h!]
	\centering
	\caption{\label{tab3} Table of the predicted radii.}
	\vspace{0.5cm}
\begin{tabular}{ccccc}
			\hline
			Compact & Observed  & Observed & Rastall & Predicted \\ 
			objects & mass $(M_{\odot})$ & radius $(Km)$ & parameter $(\xi)$ & radius $(Km)$ \\
			\hline 
			\vspace{0.1cm}
			4U 1820-30 \cite{Guver} & $1.58^{+0.06}_{-0.06}$ & $9.1^{+0.4}_{-0.4}$ & 0.01 & 9.24 \\
			\vspace{0.1cm}
			LMC X-4 \cite{Rawls} & $1.04^{+0.09}_{-0.09}$ & $8.301^{+0.2}_{-0.2}$ & 0.05 & 8.34 \\
			\vspace{0.1cm}
			Cen X-3 \cite{Rawls} & $1.49^{+0.08}_{-0.08}$ & $9.178^{+0.13}_{-0.13}$ & 0.01 & 9.10 \\
			\vspace{0.1cm}
			HER X-1 \cite{Abubekerov} & $0.85^{+0.15}_{-0.15}$ & $8.1^{+0.41}_{-0.41}$ & 0.09 & 8.12 \\
			\vspace{0.1cm} 
			4U 1608-52 \cite{Guver1} & $1.74^{+0.14}_{-0.14}$ & $9.3^{+1.0}_{-1.0}$ & 0.01 & 9.46 \\
			\hline
		\end{tabular}
\end{table}
\begin{table}[h!]
	\centering
	\caption{\label{tab4} Table of the physical parameters associated with stars given in Table~\ref{tab3}.}
		\begin{tabular}{ccccc}
			\hline
			Compact & Rastall & Central & Surface & Central\\ 
			objects & parameter $(\xi)$ & density $(gm/cm^{3})$ & density $(gm/cm^{3})$ & pressure $(dyn/cm^{2})$ \\
			\hline 
			\vspace{0.1cm}
			4U 1820-30 & 0.01 & $1.29\times10^{15}$ & $0.73\times10^{15}$ & $1.81\times10^{35}$\\
			\vspace{0.1cm}
			LMC X-4  & 0.05 & $0.89\times10^{15}$ & $0.61\times10^{15}$ & $1.05\times10^{35}$\\
			\vspace{0.1cm}
			Cen X-3 &  0.01 & $1.25\times10^{15}$ & $0.73\times10^{15}$ & $1.63\times10^{35}$ \\
			\vspace{0.1cm}
			HER X-1 & 0.09 & $0.60\times10^{15}$ & $0.44\times10^{15}$ & $0.82\times10^{35}$ \\
			\vspace{0.1cm} 
			4U 1608-52 & 0.01 & $1.37\times10^{15}$ & $0.73\times10^{15}$ & $2.20\times10^{35}$ \\
			\hline
		\end{tabular}
\end{table}  
\newpage    
\subsection{Causality conditions}
\begin{figure}[ht!]
	\centering
	\includegraphics[width=10cm]{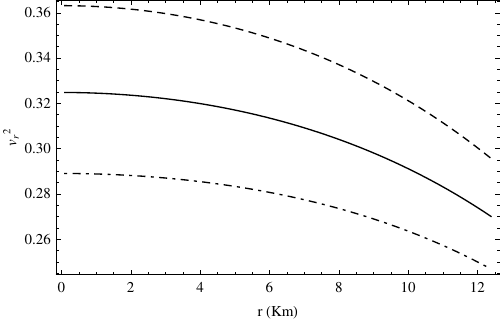}
	\caption{Variation of $ v_{r}^2$ with radius $r$. Here, Dashed, Solid and Dot-Dashed lines represent $\xi=0.01,~0.05$ and $0.09$ respectively.}
	\label{fig5}
\end{figure}

\begin{figure}[ht!]
	\centering
	\includegraphics[width=10cm]{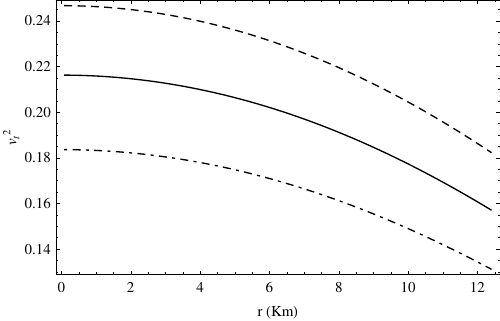}
	\caption{Variation of $ v_{t}^2$ with radius $r$. Here, Dashed, Solid and Dot-Dashed lines represent $\xi=0.01,~0.05$ and $0.09$ respectively.}
	\label{fig6}
\end{figure}

For a realistic model of an anisotropic compact star, the interior matter can be characterised by analysing the velocity of sound waves in radial and tangential directions, given by $v_{r}=\sqrt{\left(\frac{dp_{r}}{d\rho}\right)}$ and $v_{t}=\sqrt{\left(\frac{dp_{t}}{d\rho}\right)}$, where $\rho$, $p_{r}$ and $p_{t}$ have been mentioned earlier. Using the system of units where $\hbar=c=1$, the causality condition imposes an absolute upper bound on the sound velocities, requiring $v_{r}^{2}\leq1$ and $v_{t}^{2}\leq1$. Additionally, thermodynamic stability demands that $v_{r}^{2}>0$ and $v_{t}^{2}>0$. Consequently, within the stellar interior, the conditions $0<v_{r}^{2}\leq 1$ and $0<v_{t}^{2}\leq 1$ must both be satisfied. From Fig.~\ref{fig5} and Fig.~\ref{fig6}, we note that the causality condition is well maintain throughout the stellar model. 
\newpage
\subsection{Energy condition}
\begin{figure}[h]
	\centering
	\includegraphics[width=10cm]{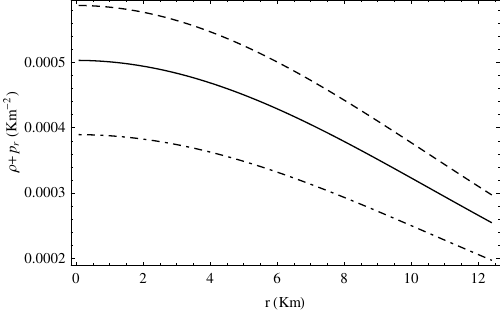}
	\caption{Variation of $(\rho+p_{r})$ with radius $r$. Here, Dashed, Solid and Dot-Dashed lines represent $\xi=0.01,~0.05$ and $0.09$ respectively.}
	\label{fig7}
\end{figure}

\begin{figure}[h]
	\centering
	\includegraphics[width=10cm]{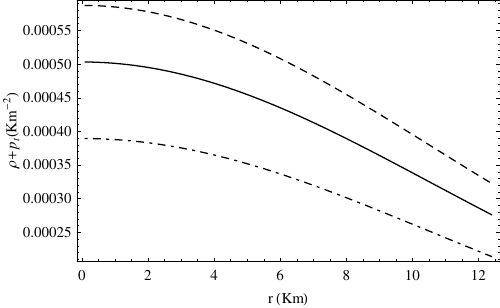}
	\caption{Variation of $(\rho+p_{t})$ with radius $r$. Here, Dashed, Solid and Dot-Dashed lines represent $\xi=0.01,~0.05$ and $0.09$ respectively.}
	\label{fig8}
\end{figure}

\begin{figure}[h]
	\centering
	\includegraphics[width=10cm]{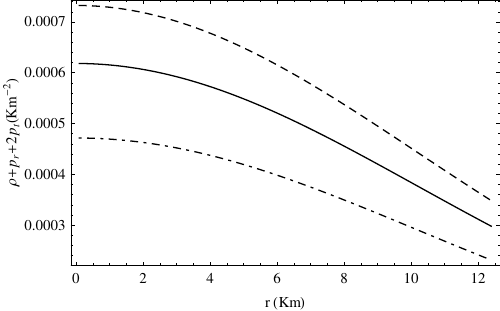}
	\caption{Variation of $(\rho+p_{r}+2p_{t})$ with radius $r$. Here, Dashed, Solid and Dot-Dashed lines represent $\xi=0.01,~0.05$ and $0.09$ respectively.}
	\label{fig9}
\end{figure}

\begin{figure}[h]
	\centering
	\includegraphics[width=10cm]{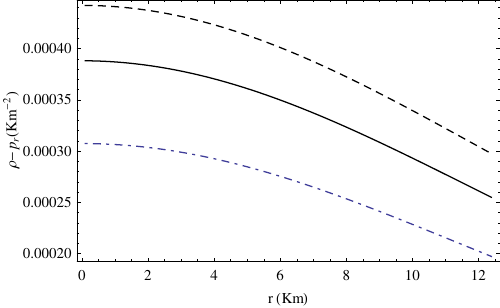}
	\caption{Variation of $(\rho-p_{r})$ with radius $r$. Here, Dashed, Solid and Dot-Dashed lines represent $\xi=0.01,~0.05$ and $0.09$ respectively.}
	\label{fig10}
\end{figure}

\begin{figure}[h]
	\centering
	\includegraphics[width=10cm]{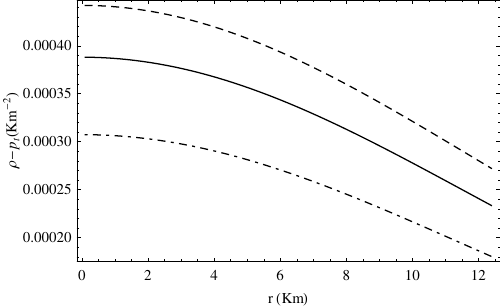}
	\caption{Variation of $(\rho-p_{t})$ with radius $r$. Here, Dashed, Solid and Dot-Dashed lines represent $\xi=0.01,~0.05$ and $0.09$ respectively.}
	\label{fig11}
\end{figure}

Energy conditions are responsible for the qualitative description of the nature of the internal matter distribution of a gravitational system. Precisely, they are the eigen value problems \cite{Kolassis} of the energy-momentum tensor. In the 4D space-time, the investigations of the matter distribution on the basis of energy conditions lead to a complex situation involving quartic polynomials which are difficult to solve due to the presence of analytical solutions. However, a perfect fluid distribution must follow the energy conditions \cite{Kolassis,Hawking,Wald} known as Null (NEC), Weak (WEC), Strong (SEC) and Dominant (DEC) energy conditions throughout the stellar boundary to emerge as a physically realistic configuration. In the case of a physically viable model of compact star, the necessary energy conditions must be fulfilled at all internal points as well as at the surface of compact stars. For the present model, we have checked the energy conditions \cite{Brassel,Brassel1} in the framework of Rastall theory of gravity within the parameter space used here. It is found that all the mentioned energy conditions hold good in Rastall theory of gravity also. In our study, we have verified the following energy conditions:\\
(i) NEC:$\rho+p_{r}\geq 0,\rho+p_{t}\geq 0$.\\
(ii) WEC:$\rho\geq 0,\rho+p_{r}\geq 0,\rho+p_{t}\geq 0$.\\
(iii) SEC: $\rho+p_{r}\geq 0,\rho+p_{t}\geq0, \rho+p_{r}+2p_{t}\geq0$.\\
(iv) DEC:$\rho\geq0,\rho-p_{r}\geq0,\rho-p_{t}\geq0$\\
The fulfillment of the above energy conditions is shown in Fig.~\ref{fig7}, Fig.~\ref{fig8}, Fig.~\ref{fig9}, Fig.~\ref{fig10} and Fig.~\ref{fig11}.
\newpage
\section{Stability analysis}\label{sec5}
The stability in Rastall theory of gravity is explored using the following methods:
\begin{enumerate}
	\item Generalised TOV equation.
	\item Cracking condition of Herrera and
	\item The study of adiabatic index of fluid
\end{enumerate}
\subsection{Generalised TOV equation}
\begin{figure}[h]
	\centering
	\includegraphics[width=10cm]{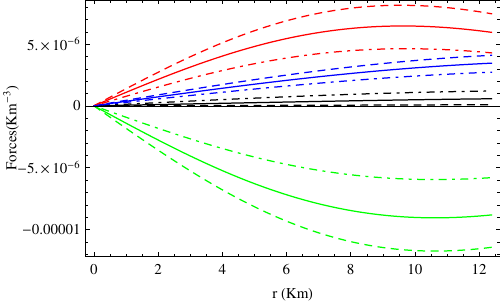}
	\caption{Variation of different forces with radius $r$. Here, Dashed, Solid and Dot-Dashed lines represent $\xi=0.01,~0.05$ and $0.09$ respectively.}
	\label{fig12}
\end{figure}

It is important to study the stability of a model under the influence of different forces. For an anisotropic compact object, the stability analysis is based on the following force components-(i) the gravitational force$(F_{g})$, (ii) the hydrostatic force$(F_{h})$ and (iii) the anisotropic force$(F_{a})$.The model should be in equilibrium under the combined influence of these forces. We have studied the stability criterion in Rastall theory of gravity using the generalised form of Tolman-Oppenheimer-Volkoff (TOV) equation \cite{Tolman,Oppenheimer} which is given below:
\begin{equation}
	-\frac{M_{G}(r)(\rho+p_{r})}{r^2}e^{\lambda -\nu}-\frac{dp_{r}}{dr}+\frac{2\Delta}{r}=0 \label{eq33}
\end{equation}
here, $M_{G}$ is termed the active gravitational mass evaluated from the Tolman-Whittaker \cite{Gron} mass formula which is given below: 
\begin{equation}
	M_{G}(r)=r^2\nu'e^{\nu-\lambda} \label{eq34}
\end{equation}
Substituting Eq.~\ref{eq34} in Eq.~\ref{eq33},we obtain
\begin{equation}
	-\nu'(\rho+p_{r})-\frac{dp_r}{dr}+\frac{2\Delta}{r}=0, \label{35}
\end{equation} 
where, $F_{g}=-\nu'(\rho +p_{r})$, $F_{h}=-\frac{dp}{dr}$ and $F_{a}=\frac{2\Delta}{r}$. In terms of the forces, $F_{g},~F_{h}$ and $F_{a}$, Eq.~\ref{eq33} can be written, in hydrostatic equilibrium, as
\begin{equation}
	F_{g}+F_{h}+F_{a}=0 \label{36}
\end{equation}
In Fig.~\ref{fig12}, we have shown the radial variation of $F_{g},~F_{h}$ and $F_{a}$ and their resultant. It is noted that the TOV equation maintains the static equilibrium condition for non-zero value of $\xi$. It is also observed that gravitational force $F_{g}$ dominates over the hydrostatic force $(F_{h})$ and anisotropic $(F_{a})$ forces. However, the sum of these forces is always zero for the present model indicating a stable stellar structure.
\subsection{Cracking condition}
Anisotropic models must remain stable when subjected to fluctuations in their physical parameters. Herrera \cite{Herrera} introduced the "cracking" condition as a method to assess the stability of these models. Building on Herrera's concept, Abreu et al. \cite{Abreu} proposed a criterion for determining the stability of an anisotropic stellar model. According to Abreu et al. \cite{Abreu}, this stability criterion is met if the squares of the radial velocity $(v_{r}^{2})$ and tangential velocity $(v_{t}^{2})$ satisfy a specific condition,
\begin{equation}
	0\leq|v_{r}^{2}-v_{t}^{2}|\leq 1, \label{eq30}
\end{equation} 
\begin{figure}[h]
	\centering
	\includegraphics[width=10cm]{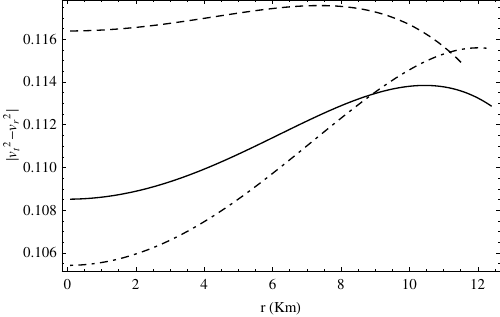}
	\caption{Variation of sound parameter $|v_{t}^2-v_{r}^2|$ with radius $r$. Here, Dashed, Solid and Dot-Dashed lines respectively represent $\xi=0.01,~0.05$ and $0.09$.}
	\label{fig13}
\end{figure}
From Fig.~\ref{fig13}, it is evident that the Abreu inequality is well satisfied in the present model.
\subsection{Adiabatic index}
\begin{figure}[h]
	\centering
	\includegraphics[width=10cm]{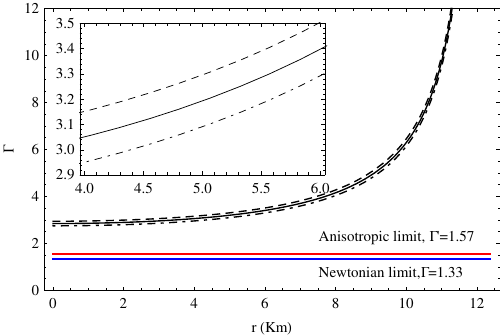}
	\caption{Variation of $\Gamma$ with radius $r$. Here, Dashed, Solid and Dot-Dashed lines represent $\xi=0.01,~0.05$ and $0.09$ respectively.}
	\label{fig14}
\end{figure}

The stiffness of EoS for a given energy density is described by adiabatic index $(\Gamma)$. It verifies the stability of stellar objects, both relativistically and non-relativistically. The adiabatic index is expressed as: 
\begin{equation}
	\Gamma=\frac{\rho+p_{r}}{p_{r}}(\frac{dp_{r}}{d\rho})=\frac{\rho+p_{r}}{p_{r}}v_{r}^2 \label{37}
\end{equation} 
According to Heintzmann and Hillebrandt \cite{Heintzmann}, the necessary condition for the stability of fluid sphere inside an isotropic star is characterised by the value of adiabatic index $(\Gamma)$ and is represented as,$\Gamma>\frac{4}{3}$ (Newtonian limit). In presence of pressure anisotropy, Chan et al. \cite{Chan} modified the expression of adiabatic index in the form:
\begin{equation}
	\Gamma'=\frac{4}{3}-\Bigg[\frac{4}{3}\frac{(p_{r}-p_{t})}{|p_{r}'|r}\Bigg]_{max}, \label{38}
\end{equation}
Now, in the Rastall theory of gravity, to ensure a stable anisotropic stellar model, we must have $\Gamma>\Gamma'$, for non-zero value of $\xi$. In Fig.~\ref{fig14}, we have shown the value of Newtonian limit, anisotropic limit $(\gamma')$ of adiabatic index and the plot of $\Gamma$ vs $r$ inside the anisotropic star for $\xi=0.01,~0.05$ and $0.09$ respectively. It is evident that potential stability is ensured for the study of adiabatic index of the fluid is Rastall theory of gravity. 
\newpage
\section{Discussion}\label{sec6}
The present article explores the structural and dynamical properties of a singularity-free anisotropic compact star within the Rastall theory of gravity framework. Considering a spherically symmetric space-time and a perfect fluid distribution, non-singular stellar solutions are obtained by solving the EFE by employing the KB metric ansatz \cite{KB}. It is well established that the additional degrees of freedom introduced by modified theories of gravity alter the boundary conditions of GR. This paper examines these adjustments by assessing the continuity of the induced metric and the extrinsic curvature tensors at the boundary of the star. In order to determine the constants of the KB metric, the interior solutions are matched to the exterior Schwarzschild vacuum solution by computing the extrinsic curvature tensors. Moreover, as part of the matching condition, we utilise the fact that the radial pressure at the surface vanishes. In this theoretical formalism, to assess the physical applicability of our model, pulsar PSR J0740+6620 (mass $2.072~M_{\odot}$ and radius 12.39 Km) is considered. We present our results through both numerical and graphical analyses. The key findings of the present model are as follows:
\begin{itemize}
	\item The radial variations of energy density as well as pressure are illustrated in Fig.~\ref{fig1}, Fig.~\ref{fig2} and Fig.~\ref{fig3} and it is noted that energy density $(\rho)$, radial $(p_{r})$ and transverse $(p_{t})$ pressures adhere to the monotonically decreasing nature from the centre to the stellar surface. Moreover, for the characteristic variation of Rastall parameter $(\xi)$, we note that as $\xi$ increases, the effective energy density and pressure profiles decrease. This may be explained as follow: with increasing $\xi$, the non-conservation of energy-momentum tensor becomes more prominent which increases the gravity-matter coupling. Consequently, with increasing magnitude of coupling, the energy density and pressure decreases. This nature is evident in Fig.~\ref{fig1}, Fig.~\ref{fig2} and Fig.~\ref{fig3}. The radial variation of anisotropy parameter $(\Delta)$ is shown in Fig.~\ref{fig4}. It is to be noted that, in the present scenario, $\Delta>0$ indicates a repulsive anisotropic nature which may be attributed to the fact that more massive stellar configuration can be explained through this formalism. 
	\item The mass-radius relation is obtained from the solution of the TOV equations. Since, TOV equations are highly dependent on the particular choice of EoS, we have evaluated the best fit EoS by the maximisation of sound velocity at the centre of the star $(r\rightarrow0)$ and are shown in Table~\ref{tab1}. The mass-radius relationship is graphically represented in Fig.~\ref{fig4a}. Within the parameter space, for the chosen Rastall parameter $(\xi)$ as 0.01, 0.05 and 0.09, we achieve a maximum mass of $2.24~M_{\odot}$, $2.28~M_{\odot}$ and $2.36~M_{\odot}$, respectively, for the corresponding radii of 9.48 Km, 9.70 Km and 10.15 Km and tabulated in Table~\ref{tab2}.
	To reiterate, as $\xi$ increases the maximum mass-radius increases, which may be attributed to the fact that with increasing $\xi$, a stellar structure's ability to counter-balance the gravitational collapse may increases, which in turn may lead to higher values of maximum mass and radius as represented in Table~\ref{tab1}. Furthermore, it must be noted that in the present model value of Rastall parameter $\xi>0.09$ gives unphysical results. Hence, we have adhered to this specific range of $\xi$. We have also predicted the radii of some recently observed compact stars and pulsars and are tabulated in Table~\ref{tab3}. we note that in the framework of Rastall theory of gravity, radii of many compact stars may be predicted, theoretically, with great accuracy by varying parameter $\xi$, so far the estimated radii of these compact stars are concerned from recent observations. 
	\item For a physically realistic stellar model, a unique way to check for the interior matter distribution as well as its stellar viability is through the investigation to sound velocity within the stellar structure which is also known as the causality condition. For a well-behaved stellar model, the condition of causality dictates that the radial $(v_{r})$ and transverse $(v_{t})$ sound velocities within the stellar boundary must be less that 1. Fig.~\ref{fig5} and Fig.~\ref{fig6} demonstrate that the causality is well satisfied in the present model. 
	\item  In our study, we have also focused on the energy conditions important in the context of theories of gravity. To simplify the analysis and avoid the intricacies of mathematical expressions, we have provided a graphical representation of the radial variation of all the energy conditions in Fig.~\ref{fig7}, Fig.~\ref{fig8}, Fig.~\ref{fig9}, Fig.~\ref{fig10} and Fig.~\ref{fig11}. The plots clearly demonstrate that our model adheres to all the required energy conditions, thereby reinforcing its validity in Rastall theory of gravity also. 
	\item We have studied the stability criterion of the present model through the analysis of generalised TOV equation, cracking condition of Herrera, and the radial variation of adiabatic index. From Fig.~\ref{fig12}, we note that our model fulfils the conditions of hydrostatic equilibrium under the combined influences of forces existing inside the stellar configuration. The stability of the stellar model on the basis of parametric fluctuations is described by the cracking condition proposed by Herrera. The stability criterion of Abreu et al. \cite{Abreu}, based on Herrera's concept, is satisfied in the present model as evident from Fig.~\ref{fig13}. Further, from Fig.~\ref{fig14}, it is evident that our model maintains the adiabatic index variation in presence of pressure anisotropy within the framework of Rastall theory of gravity. 
\end{itemize}
Therefore, based on the significant results presented in the paper, it is evident that we have successfully developed a singularity-free, physically viable, and stable stellar model within the context of Rastall theory of gravity. 
\section{Acknowledgements}
SB is thankful to the Department of Physics, Cooch Behar Panchanan Barma University, for providing the necessary help to carry out the research work. DB is thankful to the Department of Science and Technology (DST), Govt. of India for providing the fellowship vide no: DST/INSPIRE Fellowship/2021/IF210761. PKC gratefully acknowledges support from the Inter-University Centre for Astronomy and Astrophysics (IUCAA), Pune, India, under Visiting Associateship programme, where the work has been completed. DB sincerely expresses gratitude to IUCAA, Pune, for providing the visitor facilities.

\end{document}